\documentclass{iopart}
\usepackage{iopams}
\usepackage[dvips]{graphicx}
\begin{document}
%%%%%%%%%%%%%%%%%%%%%%%%%%%%%%%%%%%%%%%%%%%%%%%%%%%%%%%%%%
\newpage

\title[Bose-Einstein condensation in dense matter....]
{Bose-Einstein condensation in dense matter and the third family of compact 
stars}

\author{Sarmistha Banik and Debades Bandyopadhyay}

\address{Saha Institute of Nuclear Physics, 1/AF Bidhannagar, 
Kolkata-700064, India}

\begin{abstract}
We investigate antikaon condensation in compact star matter using a 
relativistic mean field model. Antikaon condensates make the 
equation of state softer resulting in a smaller maximum mass star compared to 
the case without condensate. It is found that the equation of state 
including antikaon condensates gives rise to a stable sequence of compact
stars called the third family beyond the neutron star branch.
\end{abstract} 
It was first demonstrated by Kaplan and Nelson \cite{Kap} within a chiral 
$SU(3)_L \times SU(3)_R$ model that $K^-$ meson may undergo Bose-Einstein 
condensation in dense matter formed in relativistic heavy ion collisions. In 
this model,
the effective mass of antikaons decreases with increasing density because of
the strongly attractive antikaon-baryon interaction. Consequently, the 
in-medium energy of $K^-$ mesons in zero momentum state also decreases with 
density. The $s$-wave $K^-$ condensation sets in when the energy of $K^-$ 
mesons equals to its chemical potential. 

Since the work of Kaplan and Nelson, there is a growing interest to understand
(anti)kaon properties in dense matter formed in relativistic heavy ion 
collisions as well as compact star matter \cite{Pal,Li,Pra,Gle1,Sami}.
The in-medium properties of (anti)kaons were studied through the analyses
of collective flow \cite{Pal} and particle spectra of (anti)kaons \cite{Li} 
and $K^-$ atomic data \cite{Fri}. All these experimental results suggest that 
kaon-nucleon interaction is repulsive whereas it is attractive for 
antikaon-nucleon interaction \cite{Pal}. Here, we investigate antikaon 
condensation in compact star matter and its role on the equation of state (EoS)
and the structure of compact stars.

\begin{figure}[t]
\begin{center}
\includegraphics[height=9cm]{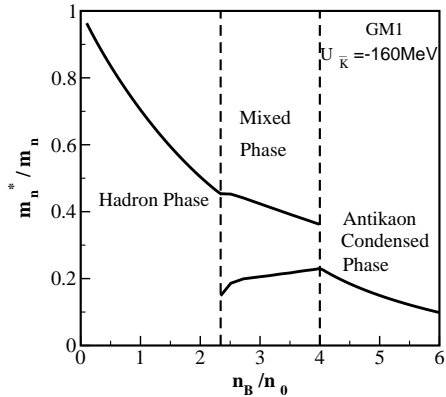} 
\caption{The effective nucleon mass is plotted with baryon number density.}
\end{center}
\end{figure}

In this calculation, $K^-$ condensation in compact star matter is treated as a
first order phase transition. We describe the equilibrium conditions of the 
pure antikaon condensed and hadronic matter and their mixed phase in a uniform
background of electrons and muons. The antikaon condensed phase is consisted of
all species of the baryon octet, leptons and antikaons. The baryon-baryon 
interaction mediated by meson exchange is described by a relativistic mean 
field model \cite{Gle1,Sami}. The model also includes hyperon-hyperon 
interaction 
through two hidden strangeness mesons $f_0(975)$ (denoted as $\sigma^*$) and 
$\phi$ (1020). The (anti)kaon-baryon Lagrangian density in the minimal coupling
scheme is ${\cal L_K} = D^*_{\mu} {\bar K} D^{\mu}K - m_K^* {\bar K} K$ with the
covariant derivative $D_{\mu} = {\partial_{\mu}} + ig_{\omega K} \omega_{\mu}
+ ig_{\phi K} \phi_{\mu} + ig_{\rho K} {\vec {t}}_K \cdot 
{\vec {\rho}}_{\mu}$. 
The effective mass of kaons 
is $m_K^* = m_K - g_{\sigma K} \sigma - g_{\sigma^* K} \sigma^*$, where
$m_K$ is the bare kaon mass. Various strangeness changing processes such as 
$n \rightleftharpoons p + K^-$ and $e^- \rightleftharpoons K^- + \nu_e$, occur
in neutron stars. It leads to the chemical equilibrium condition 
$\mu_{K^-} = \mu_e$. The onset of $K^-$ condensation is possible when
the above condition is satisfied. We also include $\bar K^0$ condensation 
in this calculation and this is treated as a second order phase transition. The
threshold condition for $\bar K^0$ condensation is $\omega_{\bar K^0} = 0$. The
constituents of this phase are in beta-equilibrium and maintain local charge
neutrality. The energy density ($\epsilon^{\bar K}$) of 
this phase has contributions from baryons, leptons 
and antikaons \cite{Gle1,Sami}. The pressure follows from the relation 
$P^{\bar K} = \sum_i{\mu_i n_i} - \epsilon^{\bar K}$, where 
$n_i$ is the number density of i-th species.

To describe the pure hadron phase, we employ the field theoretical model for
baryon-baryon interaction as described above. The equation of state of 
this phase is obtained by solving the meson field equations and effective 
baryon mass in conjunction with local charge neutrality and beta-equilibrium 
conditions \cite{Gle1,Sami}. The energy density ($\epsilon^h$) and pressure 
($P^h$) in this phase are related by $P^h = \sum_i{\mu_i n_i} - \epsilon^h$.

The mixed phase of antikaon condensed matter and hadronic matter is governed
by Gibbs phase rules and global conservation laws \cite{Glen}. The Gibbs phase 
rules read, $P^h = P^{\bar K}$ and $\mu_B^h = \mu_B^{\bar K}$ where $\mu_B^h$
and $\mu_B^{\bar K}$ are chemical potentials of baryon B in hadronic and $K^-$
condensed matter respectively. The conditions for global charge neutrality and 
baryon number conservation are $(1-\chi) Q^h + \chi Q^{\bar K}=0$ and
$n_B = (1-\chi) n_B^h + \chi n_B^{\bar K}$ where $\chi$ is the volume fraction
in the condensed phase. The total energy density in the mixed phase is given 
by $\epsilon = (1-\chi) \epsilon^h + \chi \epsilon^{\bar K}$.

\begin{figure}[t]
\begin{center}
\includegraphics[height=8cm,width=8cm]{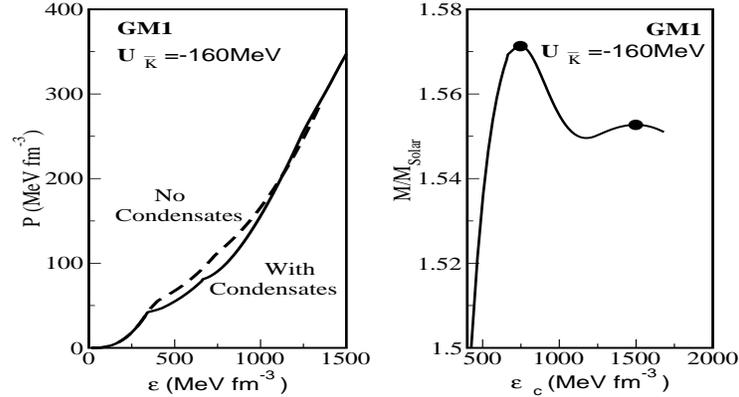} 
\caption{The equation of state with and without antikaon condensates are shown
in the left panel and the compact star mass sequences are plotted in the right 
panel.}
\end{center}
\end{figure}

In this calculation, we adopt GM1 parameter set \cite{Gle2} where nucleon-meson
coupling constants are determined from the nuclear matter saturation 
properties. The vector meson coupling constants for 
(anti)kaons and hyperons are determined from the quark model \cite{Sch1}. 
The scalar meson coupling constants for hyperons and antikaons are 
obtained from the potential depths of  hyperons and antikaons in normal nuclear 
matter \cite{Sami,Sch1}. The phenomenological fit to the $K^-$ atomic 
data yielded the real part of antikaon potential as $U_{\bar K} = -180 \pm 20$
MeV \cite{Fri}. On the other hand, the recent microscopic calculations predict 
a shallow attractive potential at the saturation density \cite{Sch2}. We 
perform this calculation with an antikaon optical potential of -160 MeV at 
normal nuclear matter density ($n_0=0.153 fm^{-3}$). The coupling 
constants for strange mesons with (anti)kaons are given as 
$g_{\sigma^* K} = 2.65$ and $\sqrt{2} g_{\phi K} = 6.04$ \cite{Sch1}. The 
coupling constants for $\sigma^*$-hyperons are calculated by fitting them to  
the potential depth for a hyperon in hyperon matter at the saturation density 
\cite{Sami,Sch1}.

The effective nucleon mass is plotted with baryon density in Figure 1. Two 
vertical dotted lines denote the lower and upper boundary of the mixed phase.
The mixed phase begins at 2.23$n_0$ and terminates at 4.0$n_0$.
The upper curve shows the behaviour of effective nucleon 
mass with density in the pure hadronic matter. With the onset of 
$K^-$ condensation, a second solution for the effective nucleon mass appears. 
This result was first obtained by Glendenning and Schaffner-Bielich \cite{Gle1}.
It shows that the effective mass of nucleons behaves differently in
the pure hadronic and antikaon condensed matter. This may be attributed to the
behaviour of $\sigma$ and $\sigma^*$ fields in those pure phases. The effective
nucleon mass in the pure antikaon condensed phase is lower 
than that of the pure hadronic phase. It is to be noted here that 
it is energetically favourable for $K^-$ mesons in the zero momentum state to 
make the system charge neutral removing electrons and muons. Just after the 
mixed phase is over, $\bar K^0$ condensation occurs at $\sim 4.1n_0$. 
It is $\Lambda$ hyperon which appears first in the mixed phase. 
Neutron and proton fractions become equal after $\bar K^0$ condensation.
With the appearance of negatively charged $\Sigma$ and $\Xi$ hyperons around
$\sim 7n_0$, the density of $K^-$ condensate starts falling. 

The equation of state for compact star matter with and without antikaon 
condensates are exhibited in the left panel of Figure 2. The curve indicating 
the overall EoS with hyperons and antikaon condensates (solid line) is softer 
compared with the EoS with hyperons and no condensate (dashed line). The kinks 
on the lower curve mark the beginning and end of the mixed phase. These kinks 
may lead to discontinuity in the velocity of sound.

Gerlach first argued that a third family of compact stars beyond white dwarf 
and neutron star branches could exist in nature \cite{Ger}. It was attributed
to the behaviour of an EoS at high density i.e. a jump in the EoS and 
consequently a discontinuity in the speed of sound and adiabatic index. 
Glendenning and Kettner
\cite{Ket} first found the actual physical situation where the EoS 
corresponding to a first order phase transition from hadronic to quark matter,
had the requisite properties. With that EoS, it was noted that the neutron star
branch was terminated due to abnormally small adiabatic index and a jump in the 
adiabatic index thereafter gave rise to another stable branch of compact stars
at higher central energy densities. Later, various groups \cite{Sami,Sche} 
obtained similar solutions in first order phase transitions from hadronic 
matter to strange matter. It is worth mentioning here that such a solution was
only possible for a subspace of the parameter space in each calculations. 
Here, we
study the structure of compact stars using Tolman-Oppenheimer-Volkoff equations
and the EoS with and without $\bar K$ condensates. The compact star mass 
sequences are shown with central energy density in the right panel of Figure 2.
For the EoS with $\bar K$ condensates, it is found from the figure that after 
the positive slope neutron star 
branch, there is an unstable region followed by another positive slope compact 
star branch. From the study of fundamental mode of radial vibration, we find 
that the third family branch is a stable one. However, there is no third family 
solution for the EoS without $\bar K$ condensate (not shown 
in the figure). The maximum masses of neutron star and third family 
branch are 1.571$M_{\odot}$ and 1.553$M_{\odot}$ corresponding to radii 12.8 km
and 10.7 km. Because of partial overlapping mass regions of the neutron star
and third family branch, nonidentical stars having same mass but distinctly
different composition and radii could exist. Such a pair of compact stars is
called "neutron star twins" \cite{Ket}. It would be challenging to observe 
two stars with almost the same mass but different radii as the proof of the 
existence of a third family solution.


\section*{References}

\begin{thebibliography}{99}

\bibitem{Kap} Kaplan D B and Nelson A E 1986 {\it Phys. Lett.} B {\bf 175} 57\\
Nelson A E and Kaplan D B 1987 {\it ibid} {\bf 192} 193   
\bibitem{Pal} Pal S, Ko C M, Lin Z and Zhang B 2000 {\it Phys. Rev.} C 
{\bf 62} 061903(R)
\bibitem{Li} Li G Q, Lee C -H, and Brown G E 1997 {\it Phys. Rev. Lett.} 
{\bf 79} 5214
\bibitem{Pra} Prakash M, Bombaci I, Prakash M, Ellis P J, Lattimer J M and 
Knorren R 1997 {\it Phys. Rep.} {\bf 280} 1   
\bibitem{Gle1} Glendenning N K and Schaffner-Bielich J 
1998 {\it Phys. Rev. Lett.} {\bf 81} 4564\\
Glendenning N K and Schaffner-Bielich J 
1999 {\it Phys. Rev.} C {\bf 60} 025803
\bibitem{Sami} Banik S and Bandyopadhyay D 
2001 {\it Phys. Rev.} C {\bf 64} 055805
\bibitem{Fri} Friedman E, Gal A, Mare\v{s} J and Ciepl\'y A
1999 {\it Phys. Rev.} C {\bf 60} 024314
\bibitem{Glen} Glendenning N K
1992 {\it Phys. Rev.} D {\bf 46} 1274
\bibitem{Gle2} Glendenning N K and Moszkowski S A
1991 {\it Phys. Rev. Lett.} {\bf 67} 2414
\bibitem{Sch1} Schaffner J and Mishustin I N
1996 {\it Phys. Rev.} C {\bf 53} 1416
\bibitem{Sch2} Schaffner-Bielich J, Koch V and Effenberger M
2000 {\it Nucl. Phys.} A {\bf 669} 153
\bibitem{Ger} Gerlach U H 1968 {\it Phys. Rev.} {\bf 172} 1325
\bibitem{Ket} Glendenning N K and Kettner C
2000 {\it Astron. Astrophys.} {\bf 353} L9
\bibitem{Sche} Schertler K, Greiner C, Schaffner-Bielich J and Thoma M H
2000 {\it Nucl. Phys.} A {\bf 677} 463\\
Fraga E S, Pisarski R D and Schaffner-Bielich J
2001 {\it Phys. Rev.} D {\bf 63} 121702\\
Schaffner-Bielich J, Hanauske M, St\"ocker H and Greiner W
2000 {\it astro-ph/0005490}
\end{thebibliography}
\end{document}